\documentclass[
superscriptaddress,dvips
prb,twocolumn
]{revtex4-1}

\usepackage{graphicx,amsmath,amssymb,subfigure}
\usepackage{dcolumn}
\usepackage{bm}
\usepackage{color}
\begin{document}

\preprint{APS/123-QED}

\title{Exact solution and high temperature series expansion study of the 1/5-th depleted square lattice Ising model}

\author{Simeon Hanks}
\affiliation{Department of Chemistry and Physics, Georgia Regents University, 2500 Walton Way, Augusta, GA 30904}
\author{Trinanjan Datta}
\email[Corresponding author:]{tdatta@gru.edu}
\affiliation{Department of Chemistry and Physics, Georgia Regents University, 2500 Walton Way, Augusta, GA 30904}
\author{Jaan Oitmaa}
\email[Corresponding author:]{j.oitmaa@unsw.edu.au}
\affiliation{School of Physics, University of New South Wales, Sydney 2052, Australia}

\date{\today}

\begin{abstract} 
The critical behavior of the 1/5-depleted square-lattice Ising model with nearest neighbor ferromagnetic interaction has been investigated by means of both an exact solution and a high-temperature series expansion study of the zero-field susceptibility.  For the exact solution we employ a decoration transformation followed by a mapping to a staggered 8-vertex model. This yields a quartic equation for the critical coupling giving $K_{c} (\equiv\beta J_{c}) =0.695$. The series expansion for the susceptibility, to $\mathcal{O}(K^{18})$, when analyzed via standard Pad\'{e} approximant methods gives an estimate of K$_{c}$, consistent with the exact solution result to at least four significant figures. The series expansion is also analyzed for the leading amplitude and subdominant terms. 
\begin{description}
\item[PACS numbers] 0.50.50.+q, 75.10.Hk, 05.70.Jk, 64.60.fd
\end{description}
\end{abstract}
\maketitle
\section{Introduction\label{sec:intro}}
\begin{figure}[t]
\centering
\includegraphics[width=2.5in]{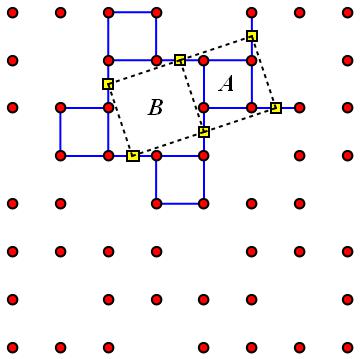}
\caption{\label{fig:deplattice} (Color online) 1/5-th depleted lattice Ising model. Filled red circles denote points on the original lattice. Yellow squares denote points on the decorated lattice. Solid blue lines denote bonds on the original lattice. Dashed black lines denote bonds on the decorated lattice. \textit{A} and \textit{B} represent the two types of plaquettes created after the decoration transformation. The transformed lattice has a staggered configuration of \textit{A} and \textit{B} plaquettes.}
\end{figure}

Exact solutions of lattice models play an important role in the study of phase transition and critical phenomena. In his seminal work, Onsager solved the two dimensional (2D) square lattice Ising model (2D-Ising) exactly \cite{onsagerPR.65.117}. Solutions have been obtained for other regular 2D lattices \cite{greenhurst}. A number of complex configurations such as the Union Jack, the bathroom tile (or 4-8),  the 4-6, and the 1/9$^{th}$ depleted lattice models have also been investigated \cite{vaksZETF.49.1180,linwangPLA.125.3,choybaxterPLA.125.365,utiyamaPTP.6.907,
linetalPLA.121.443,choybaxterJPA.21.2143}.  The 1/5-th depleted antiferromagnetic S=1/2 Heisenberg model has been an earlier topic of investigation \cite{zhengetalPRB.55.11377}. In this article we predict the critical point of a previously unexplored 1/5-th depleted Ising model on a square lattice. The Ising model Hamiltonian is given by
\begin{equation}
\mathcal{H}=-J\sum_{\langle i,j\rangle}s_{i}s_{j} -h\sum_{i} s_{i},
\label{eq:depising}
\end{equation}
where s$_{i}$ is the classical dimensionless Ising variable at site i taking the values $\pm$ 1. J $>$ 0 (ferromagnetic interaction) and h (magnetic field) denote constant parameters with dimensions of energy. The structure of the depleted lattice with every 5$^{th}$ missing site is shown in Fig.~\ref{fig:deplattice}. 

In principle all planar Ising models (i.e. with non-crossing bonds) are solvable by the Pfaffian method \cite{greenhurst}.  The method has been utilized to solve a variety of lattice models including the 8-vertex model. The 8-vertex model has been investigated both for translationally invariant and staggered vertex weight \cite{wulinPRB.12.419,wulinPRB.12.429}. In the staggered model the vertex weights are allowed to vary taking different values on the staggered plaquettes of the square lattice. The relevance of the staggered model lies in its relationship to a number of important models in statistical mechanics - the percolation model \cite{temperleyliebPRSLA.322.251}, the Potts model \cite{baxterJPC.6.L445}, and the Ashkin-Teller model \cite{wegnerJPC.5.L131,wulinJPC.9.L181}.

The partition sum of the 2D-Ising model has been evaluated using a variety of techniques [REF]. We compute the partition sum to obtain the critical point of the 1/5-th depleted Ising model with two methods. First, we carry out an exact solution by using a decoration transformation \cite{fisherPR.113.969} followed by a mapping to a staggered 8-vertex model \cite{wulinPRB.12.419,wulinPRB.12.429}. Second, we obtain the high temperature series expansion (HTSE) for the zero-field susceptibility upto $\mathcal{O}(K^{18})$ where $K=\beta J$ ($\beta=1/k_{B}T$). T is the temperature and k$_{B}$ the Boltzmann constant. Using Pad\'{e} approximants (PA) we analyze the series for its leading amplitude and subdominant terms. 

This paper is organized as follows. In Section \ref{sec:intro} we introduce the 1/5-th depleted lattice Ising model. In Section \ref{sec:exact} we obtain the exact solution. In Section \ref{sec:results} we compute the HTSE for the zero-field susceptibility and display the series coefficient results. In Section \ref{sec:analysis} we analyze the leading and subleading amplitudes of our series expansion result. Finally in Section \ref{sec:sumcon}  we summarize and conclude the main results of the paper. 
{\section{Exact solution}\label{sec:exact}}
\begin{figure}[t]
\centering
\includegraphics[width=3.0in]{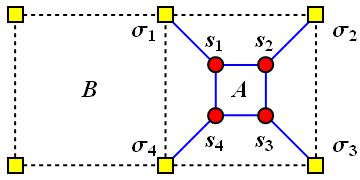}
\caption{\label{fig:plaquette} (Color online) Ising spins on the original lattice are denoted by s$_{i}$, \textit{i}=1,2,3,4 (filled red circles). Spins on the transformed decorated lattice are given by $\sigma_{i}$, \textit{i}=1,2,3,4 (yellow squares). Interactions between s-spins is given by $\mathit{K}$. Interactions between s and $\sigma$-spins is given by $\mathit{K^{'}}$.}
\end{figure}
We carry out a two step procedure to obtain the exact solution. In the first step a decoration transformation is performed. A new spin $\sigma$ (filled yellow squares in Fig.~\ref{fig:deplattice}) is introduced in the original lattice placed at the midpoint of bonds which do not lie in the small square (see Fig.~\ref{fig:deplattice}) \cite{fisherPR.113.969}. The interactions on these bonds are replaced by new interactions with the $\sigma$ spins to obtain the relation between the spins (s) in the original lattice and the decorated lattice ($\sigma$). The scaled interactions $\mathit{K}$ and $\mathit{K^{'}}$ are defined by
\begin{equation}
e^{K s_{1}s_{2}}=A\sum_{\{\sigma\}}e^{K^{'}\sigma s_{1}}e^{K^{'}\sigma s_{2}}.
\end{equation}
Carrying out the summation we obtain the following relation between $\mathit{K}$ and $\mathit{K^{'}}$
\begin{equation}
e^{2K}=\cosh(2K^{'}).
\label{eq:decotrans}
\end{equation}
The decoration process introduces two different types of plaquettes \textit{A} and \textit{B} as shown in Fig.~\ref{fig:plaquette}. We map this configuration to a staggered 8-vertex model. This transforms the 1/5-th depleted Ising model to a square lattice with more complex interactions with nearest neighbour coupling $\tilde{K}$, a diagonal next-nearest-neighbour coupling $ \tilde{L}$, and a four-spin coupling $\tilde{M}$ \cite{oitmaakeppertJPA.35.2002}. The transformed Z is 
\begin{eqnarray}
Z(\sigma_{1},\sigma_{2},\sigma_{3},\sigma_{4})=\sum_{\{s\}}\exp[K^{'}(s_{1}\sigma_{1}+s_{2}\sigma_{2}+s_{3}\sigma_{3}+s_{4}\sigma_{4})]\nonumber\\
\times\exp[K(s_{1}s_{2}+s_{2}s_{3}+s_{3}s_{4}+s_{1}s_{1})].\nonumber\\
\label{eq:origzeightvertex}
\end{eqnarray}
Enumerating the summation over the original lattice spins, s$_{i}$ (i=1,2,3,4), we can write
\begin{eqnarray}
Z(\sigma_{1},\sigma_{2},\sigma_{3},\sigma_{4})=\Lambda(K)\exp[\tilde{K}(\sigma_{1}\sigma_{2}+\sigma_{2}\sigma_{3}+\sigma_{3}\sigma_{4}+\sigma_{1}\sigma_{4})]\nonumber\\
\times\exp[\tilde{L}(\sigma_{1}\sigma_{3}+\sigma_{2}\sigma_{4})]\exp[\tilde{M}(\sigma_{1}\sigma_{2}\sigma_{3}\sigma_{4})],\nonumber\\
\end{eqnarray}
\begin{widetext}
with
\begin{eqnarray}
&\Lambda e^{4\tilde{K}+2\tilde{L}+\tilde{M}}=4 e^{-2 K^{'}}+4 e^{2 K^{'}}+e^{4 K-4 K^{'}}+e^{4 K^{'}+4 K}+2 e^{-4 K}+4=P_{1}(K), \\
&\Lambda e^{-4\tilde{K}+2\tilde{L}+\tilde{M}}=4 e^{-2 K^{'}}+4 e^{2 K^{'}}+e^{-4 K^{'}-4 K}+e^{4 K^{'}-4 K}+2 e^{4 K}+4=P_{2}(K),\\
&\Lambda e^{-2\tilde{L}+\tilde{M}}=e^{-4 K^{'}}+4 e^{-2 K^{'}}+4 e^{2 K^{'}}+e^{4 K^{'}}+2 e^{-4 K}+2 e^{4 K}+2=P_{3}(K),\\
&\Lambda e^{-\tilde{M}}=e^{-4 K^{'}}+2 e^{-2 K^{'}}+2 e^{2 K^{'}}+e^{4 K^{'}}+e^{-2 K^{'}-4 K}+e^{4 K-2 K^{'}}+e^{2 K^{'}-4 K}+e^{2 K^{'}+4 K}+6=P_{4}(K).
\end{eqnarray}
\end{widetext}
The vertex weights satisfy the free fermion condition \cite{wulinPRB.12.429}. Using Eqs.~13 and 21 from Ref.~\onlinecite{wulinPRB.12.429} the condition for the critical point for our model is 
\begin{equation}
P_{1}(K)+P_{2}(K)-2P_{3}(K)-4 P_{4}(K)=0.
\end{equation}
The above equation can be written in terms of the variable $x=e^{2K}$ as
\begin{equation}
x^{4} - 4 x^{3} - 1=0.
\label{eq:xe2k}
\end{equation}
Solving the above equation we obtain a physical root at $x = 4.015445$, yielding a critical coupling value of $K_{c} = 0.695074$. In terms of the variable $v = \tanh(K)$ Eq.~\ref{eq:xe2k} takes the form
\begin{equation}
v^4 + 4 v^3 - 1=0,
\label{eq:eqk}
\end{equation}
giving $v_{c}$=0.601232 as the solution of the critical point. 
{\section{High Temperature Series}\label{sec:results}} 
The HTSE technique is one of the most effective approaches to study critical phenomena ~\cite{dombgreen}. Much work has been devoted to the HTSE of the Ising model \cite{buteraperniciPRE.86.011139,buteracomiPRB.65.144431,hellmundwolfhardPRB.74.144201}. 
\begin{table}[h]
\caption{\label{tab:coeff} High-temperature series expansion coefficients, a$_{r}$, of the zero-field susceptibility for the 1/5 depleted Ising model. The expansion parameter is $\mathit{v}$ = tanh($\beta$J).}
\begin{ruledtabular}
\begin{tabular}{cc}
Order    & Coefficient \\ \hline
 1          &0.300000000000D+01\\
 2          &0.600000000000D+01\\
 3          &0.120000000000D+02\\
 4          &0.220000000000D+02\\
 5          &0.400000000000D+02\\
 6          &0.740000000000D+02\\
 7          &0.136000000000D+03\\
 8          &0.246000000000D+03\\
 9          &0.444000000000D+03\\
10         &0.782000000000D+03\\
11         &0.137200000000D+04\\
12         &0.240600000000D+04\\
13         &0.420800000000D+04\\
14         &0.738600000000D+04\\
15         &0.129240000000D+05\\
16         &0.223940000000D+05\\
17         &0.387280000000D+05\\
18         &0.667820000000D+05
    \end{tabular}
\end{ruledtabular}
\end{table}
Thermodynamic properties are derivable from the partition function
\begin{eqnarray}
\mathcal{Z}&=&\sum_{{s}}\exp(-\beta \mathcal{H}),\\
                   &=&\sum_{{s}}\exp(K\sum_{\langle i,j\rangle}s_{i}s_{j} +\beta h\sum_{i} s_{i}),
\label{eq:partition}
\end{eqnarray}
The zero field susceptibility, $\chi$, is given by
\begin{equation}
\chi(\textit{v})=\beta^{-1}\lim_{h\rightarrow 0}\frac{\partial^{2}}{\partial h^{2}}\left (\frac{1}{N}\ln \mathcal{Z}\right).
\label{eq:suscep}
\end{equation}
Using the identity below for both the regular and the field term
\begin{equation}
\exp(K s_{i}s_{j})= \cosh K(1+\textit{v}s_{i}s_{j}),
\label{eq:partition}
\end{equation}
we can construct a graphical expansion. Each bond carries a factor of \textit{v}s$_{i}$s$_{j}$ and, in addition, each site has a factor of either 1 or $\tau s_{k}$. Only those graphs with precisely two factors of $\tau s_{k}$ contribute to the above equations. As a result the graphs which contribute are those with precisely two vertices of odd degree, those to be compenstated by the two $\tau s_{k}$ factors. We then obtain the following result
\begin{eqnarray}
\frac{\mathcal{Z}}{(\cosh K)^{2N}(\cosh \beta h)^{N}}&=&\sum_{\{s\}}\prod_{\langle ij\rangle}(1+vs_{i}s_{j})\prod_{k}(1+\tau s_{k}) \nonumber\\
\label{eq:magpart}
\end{eqnarray}
where $v=\tanh \beta J$ and $\tau=\tanh \beta h$. The high temperature susceptibility can be expanded in the form
\begin{equation}
\beta^{-1}\chi(v)=1+\sum_{r=1}^{\infty}a_{r}v^{r}
\label{eq:chihight}
\end{equation}
The coefficients a$_{r}$ can be related to the graph counting problem and evaluated exactly \cite{dombgreen}. The computed series expansion coefficients for the zero-field susceptibility of the 1/5-th depleted Ising model are listed in Table \ref{tab:coeff}.
{\section{Susceptibility analysis}\label{sec:analysis}} 
\begin{table}[b]
\caption{\label{tab:critweight} Pade approximation analysis of the high-temperature series expansion coefficients of the zero field susceptibility for the 1/5 depleted Ising model. The critical coupling constant and the leading amplitude is listed below.}
\begin{ruledtabular}
\begin{tabular}{cccc}
(N,D)     & $v_{c}$       & A$_{0}$       & A$_{1}$\\ \hline
 (10,8)   &0.601461      & 0.686077     & 0.707575\\
 (9,9)     &0.601405      & 0.686598     & 0.707919\\
 (8,10)   &0.601437      & 0.686265     & 0.707625\\
 (9,8)     &0.601554      & 0.686771      & 0.707614\\
 (8,9)     &0.600934      & 0.686810        & 0.707385\\ 
\end{tabular}
\end{ruledtabular}
\end{table}
The universality hypothesis in critical phenomena implies that thermodynamic quantities are not sensitive to the microscopic details of a system near a critical point \cite{orrickPRL.86.4120,gartenhausscottPRB.38.11688}. It is known from earlier work that near the transition point the high temperature susceptibility, $\chi(v)$ , of the 2D-Ising model on all 2D lattices has an asymptotic form. For our model we can express the susceptibility as 
\begin{equation}
\chi(v)=A_{o}\left(1-\frac{v}{v_{c}}\right)^{-7/4}+A_{1}\left(1-\frac{v}{v_{c}}\right)^{-3/4}+\cdots
\end{equation}
with $v_c$=0.601232. To analyze the $\chi(v)$ series for its pole and its leading and subleading amplitude we first consider constructing the series
\begin{equation}
f_{1}(v)=[\chi(v)]^{4/7} \sim  A^{4/7}_{o}(1-v/v_{c})^{-1}+\cdots
\label{eq:f1} 
\end{equation}
Direct PA's to $f_{1}(v)$ give a consistent pole at $v_{c}$ $\sim$ 0.6015 $\pm$ 0.0002. This result is close to the exact solution value of $v_c$=0.601232. The residues, which are estimates of A$^{4/7}_{o}v_{c}$ are all in the range 0.486 - 0.490. Considering the value to be 0.488 we obtain A$_{o}$ $\sim$ 0.694. A more consistent set of results can be obtained by constructing the series 
\begin{eqnarray}
&f_{2}(v)=(1-v/v_{c})^{7/4}\chi(v)\nonumber\\ \sim  & A_{o}
             +\text{terms which vanish at v$_{c}$}
\label{eq:f2} 
\end{eqnarray}
and forming PA's to $f_{2}(v)$. Evaluating these at $v_c$=0.601232 gives A$_{o} \sim$ 0.687 $\pm$ 0.001. To obtain the subdominant contribution we analyze the function
\begin{eqnarray}
&f_{3}(v)=(1-v/v_{c})^{3/4}\chi(v)f_{2}(v)\nonumber\\ \sim  & A_{1}
             +\text{terms which vanish at v$_{c}$}.
\label{eq:f2} 
\end{eqnarray}
PA's of $f_{3}(v)$ computed at $v_c$=0.601232 provide a consistent set of estimates for A$_{1} \sim$ 0.708 $\pm$ 0.001. 

The above analysis can be repeated with the HTSE $\chi$ series expressed in the K variable,
\begin{equation}
\chi(K)=C_{o}\left(1-\frac{K}{K_{c}}\right)^{-7/4}+C_{1}\left(1-\frac{K}{K_{c}}\right)^{-3/4}+\cdots.
\end{equation}
To  obtain a consistent set of critical coupling value, K$_{ c}$ ,  we perform the PA analysis on a $\chi^{4/7}$ series. The leading amplitude, C$_{o}$, can be computed by investigating the PA analysis of $\left(1-K/K_{c}\right)^{7/4}\chi$. However, such an analysis does not lead to a consistent set of values for $C_{1}$. We therefore obtain both $C_{0}$ and $C_{1}$ from $A_{0}$ and $A_{1}$. To do so we expand the $\chi(v)$ series in a Taylor series in 1- $K/K_c$ up to 2nd order to obtain the following
\begin{eqnarray} 
K_{c}&=&0.695\pm 0.001, \\
C_{0}&=&1.167\pm 0.001, C_{1}=0.036\pm 0.001.
\end{eqnarray}
The agreement of the series estimate of K$_c$ with our result from Eq.~\ref{eq:xe2k} provides confirmation that our analysis in Section ~\ref{sec:exact} is correct.
{\section{Conclusion}\label{sec:sumcon}} 
We have obtained the critical point exactly, and the estimated values of the leading two amplitudes of the asymptotic form of the zero-field susceptibility for the Ising model on an unusual lattice obtained by regularly removing the 1/5-th of the sites of a square lattice. There has not been to our knowledge, any systematic study of Ising models on depleted lattices. It is worth noting, however, that the familiar honeycomb and kagome lattices result from particular 1/3-rd and 1/4-th depletions of the triangular lattice. The 1/5-th depleted square lattice, considered here, is in fact realized in the material CaV$_{4}$O$_{9}$, but not as an Ising system.

To obtain the critical point, we relate partition function of the model to that of a staggered eight-vertex model and use established results for that model. The vertex weights satisfy a `free fermion' condition, confirming that the model lies in the normal Ising universality class. We note that in the 1/5-th depleted lattice there are two classes of nearest-neighbor bonds. While we have only considered the case of equal strengths, our transformation method applies equally well to the more general case of different couplings J, J$^{'}$.

We have also derived an 18-term high temperature series for the zero-field susceptibility. Because of the low coordination number and open structure of the lattice, the number of graphs that contribute at a given order is much reduced. As a consequence, the series is not as well behaved as that of the parent square lattice. However, using standard PA methods, we obtain an estimate of the critical temperature in good agreement with the exact value. As explained in Section~\ref{sec:analysis} we obtain rather precise estimate of the leading two amplitudes in the asymptotic form of $\chi(v)$ near the critical point and, rather less precise, estimates of the amplitudes in the K-representation. 

For the square lattice exact expressions of the spin-spin correlation functions allow these amplitudes to be obtained, essentially exactly, from the solution of a Painlev\'{e} equation III ~\cite{barouchPRL.31.1409}. 
In addition an exact result relating the coefficients $C_{0}$, $C_{1}$, viz., $C_{1}/C_{0}$ =$\sqrt{2}K_{c}$/8, has been proven. It is not clear whether a similar calculation could be done for the depleted lattice. However, the ratio $C_{1}/C_{0}$, in this case, does not appear to satisfy a simple relationship of the above type. 

Finally, we remark that depletion of any regular lattice will reduce the average coordination number and this leads to a \emph{less rigid} structure. Hence the ordered state will be less robust to thermal fluctuations, and the critical temperature will be lowered. This is seen in our study, with $k_{B}T_{c}/J$ being reduced by some 36\%, from 2.2692$\dots$ to 1.4387$\dots$.
\begin{acknowledgments}
Simeon Hanks acknowledges financial support from the Savannah River Scholars Program (NSF-DUE Grant \# 0966195). T.D. acknowledges Cottrell Research Corporation Grant \# 20073 and Georgia Regents University College of Sciences and Mathematics for funding support. 
\end{acknowledgments}
\bibliography{seref}
\end{document}